\documentclass[prl,noshowpacs,english,superscriptaddress,twocolumn]{revtex4}
\usepackage{amsmath,amssymb,amsfonts,mathrsfs}
\usepackage{amsthm}
\usepackage{CJK}
\usepackage{bm}
\usepackage{babel}
\usepackage{graphicx}
\allowdisplaybreaks
\usepackage{braket}
\usepackage[breaklinks]{hyperref}
\usepackage{color}
\hypersetup{colorlinks=true, linkcolor=blue, citecolor=blue, filecolor=blue, urlcolor=blue}

\begin{document}

\title{Symmetry-Protected Ideal Type-II Weyl Phonons in CdTe}

\author{B. W. Xia}
\affiliation{Department of Physics $\&$ Shenzhen Key Laboratory of Quantum Science and Engineering, Southern University of Science and Technology,
Shenzhen 518055, P. R. China.}
\affiliation{Department of Physics, The Hong Kong University of Science and Technology,Clear Water Bay, Hong Kong, P. R. China.}

\author{R. Wang}
\affiliation{Department of Physics $\&$ Shenzhen Key Laboratory of Quantum Science and Engineering, Southern University of Science and Technology,
Shenzhen 518055, P. R. China.}
\affiliation{Institute for Structure and Function $\&$
Department of physics, Chongqing University, Chongqing 400044, P. R. China.}

\author{Z. J. Chen}
\affiliation{Department of Physics $\&$ Shenzhen Key Laboratory of Quantum Science and Engineering, Southern University of Science and Technology,
Shenzhen 518055, P. R. China.}
\affiliation{Department of Physics, South China University of Technology, Guangzhou 510640, P. R. China.}

\author{Y. J. Zhao}
\affiliation{Department of Physics, South China University of Technology, Guangzhou 510640, P. R. China.}

\author{H. Xu}
\email[]{xuh@sustc.edu.cn}
\affiliation{Department of Physics $\&$ Shenzhen Key Laboratory of Quantum Science and Engineering, Southern University of Science and Technology,
Shenzhen 518055, P. R. China.}
\affiliation{Center for Quantum Computing, Peng Cheng Laboratory, Shenzhen 518055, China.}

\begin{abstract}
Nontrivial low-energy excitations of crystalline solids have insightfully strengthened understanding of elementary particles in quantum field theory. Usually, topological quasiparticles are mainly focused on fermions in topological semimetals. In this work, we alternatively show by first-principles calculations and symmetry analysis that ideal type-II Weyl phonons are present in zinc-blende cadmium telluride (CdTe), a well-known II-VI semiconductor. Importantly, these type-II Weyl phonons originate from the inversion between the longitudinal acoustic and transverse optical branches.  Symmetry guarantees the type-II Weyl points to lie along the high-symmetry lines at the boundaries of Brillouin zone even with breaking the inversion symmetry, exhibiting the robustness of protected phonon features. The nontrivial phonon surface states and surface arcs projected on the semi-finite (001) and (111) surfaces are investigated. The phonon surface arcs connecting the Weyl points with opposite chirality, guaranteed to be very long, are clearly visible. This work not only offers a promising candidate for studying type-II Weyl phonons, but also provides a route to realize symmetry-protected nontrivial phonons and related applications in realistic materials.
\end{abstract}



\maketitle
Elementary particles such as fermions and bosons play fundamental roles in our study of the universe \cite{braibant2011particles}. There are three types of fermions in the quantum field theory, i.e., Dirac, Weyl, and Majorana fermions. Dirac fermions have been known over the past decades, while the latter two have never been observed as elementary particles in experiments of high-energy physics. Recently, condensed-matter systems have provided fascinating avenues for the discovery of fermions as quasiparticle excitations in realistic materials, such as Majorana excitations in topological superconductors \cite{PhysRevLett.100.096407, ZSC-RevModPhys.83.1057, Nadj-Perge602} and Dirac or Weyl excitations in topological semimetals \cite{Weng2015, Wang2016prl1, PhysRevB.84.235126, RevModPhys.90.015001}. These topological quasiparticles immensely strengthen understanding of elementary particles. More importantly, topological semimetals with nontrivial band structures are characterized by topological invariants, extending the topological classification of matter beyond insulators \cite{RevModPhys.88.021004, Kane-RevModPhys.82.3045}. For these topological semimetallic states, exotic features such as Fermi arc surface states and unusual transport properties strongly stimulate fundamental and practical interests due to their potential applications. As is well-known, quasiparticle excitations in crystalline solids are constrained by the crystal symmetry of space group rather than the Poincare symmetry. Therefore, unconventional topological quasiparticles without high-energy physics counterparts occur in condensed matter systems \cite{Bradlynaaf5037, lvNature2017, ZhuPhysRevX.6.031003, NatureSoluyanov2015}. For instance, there are two distinct types of Weyl fermions in Weyl semimetals. The conventional type-I Weyl fermions are associated with a closed pointlike Fermi surface. In contrast, the unconventional type-II Weyl fermions leave an open Fermi surface and don't satisfy the Lorentz symmetry in the standard model, resulting in anisotropic chiral anomaly \cite{NatureSoluyanov2015, Wang2016prl2, Autes2016}.

Overall, the advancements of topologically protected fermions in crystalline solids provide promising avenues for investigations of elementary particles. Recently, beyond nontrivial fermionic excitations, other elementary excitations such as topological bosons have been intensively studied, e.g., electromagnetic waves in photonic crystals \cite{PhysRevLett.100.013904, Naturephotonic2013, PhysRevLett.100.013905} and classical elastic waves in  phononic crystals \cite{Natcom2015, Science2015, Natphys2018}. Besides these nontrivial bosons in artificial periodic structures, intrinsic quantized collective excitations of atomic vibrations at THz frequency (i.e., topological phonons in crystalline solids) are of particular importance \cite{NCR}, which can promote fundamental investigations and promising applications related to phonons, such as dissipationless phonon transport, quantized phonon Hall effects \cite{PhysRevLett.105.225901, PhysRevLett.113.265901}, and topological thermal devices \cite{PhysRevB.96.064106}. Similar to nontrivial fermionic electrons, there are also unconventional topological phonons in crystalline solids because phonons are constrained by crystal symmetries. For instance, Weyl, double Weyl, and triple phonons in three-dimensional compounds have been recently proposed \cite{PhysRevB.97.054305, PhysRevLett.120.016401}. However, for topological phonons studied to date, these protected phonons in realistic materials is very limited. In particular, candidates with individual type-II Weyl phonons have not been reported. To intuitively exhibit the topological features of type-II Weyl phonons, it is highly desirable to explore synthesized candidates in which only ideal type-II Weyl phonons are present.

In this work, we show by first-principles calculations and an effective model analysis  that ideal type-II Weyl phonons are identified in zinc-blende cadmium telluride (CdTe) with the space group $F$-$43m$ (No. 216). These type-II Weyl phonons are formed due to the fact of phonon branch inversion between the longitudinal acoustic and transverse optical branches. In contrast to fermionic electrons, the symmetry-protected Weyl points (WPs) in CdTe are explicitly located along the  high-symmetry lines due to the absence of spin-orbit coupling (SOC) in phonon systems. We reveal that the phonon WPs with opposite chirality are well separated in momentum space, and this far separation guarantees long nontrivial phonon surface arcs. More importantly, CdTe is a well-known synthesized II-VI semiconductor \cite{ref1} whose lattice dynamics as well as phonon spectra have been widely studied both in theories and experiments \cite {Rowe1974, PhysRevB.56.8691, PhysRevB.53.9052, PhysRevB.47.3588}, implying that CdTe is an ideal and realistic candidate for utilizing its nontrivial phonon features in practical applications.

\begin{figure}
	\centering
	\includegraphics[scale=0.42]{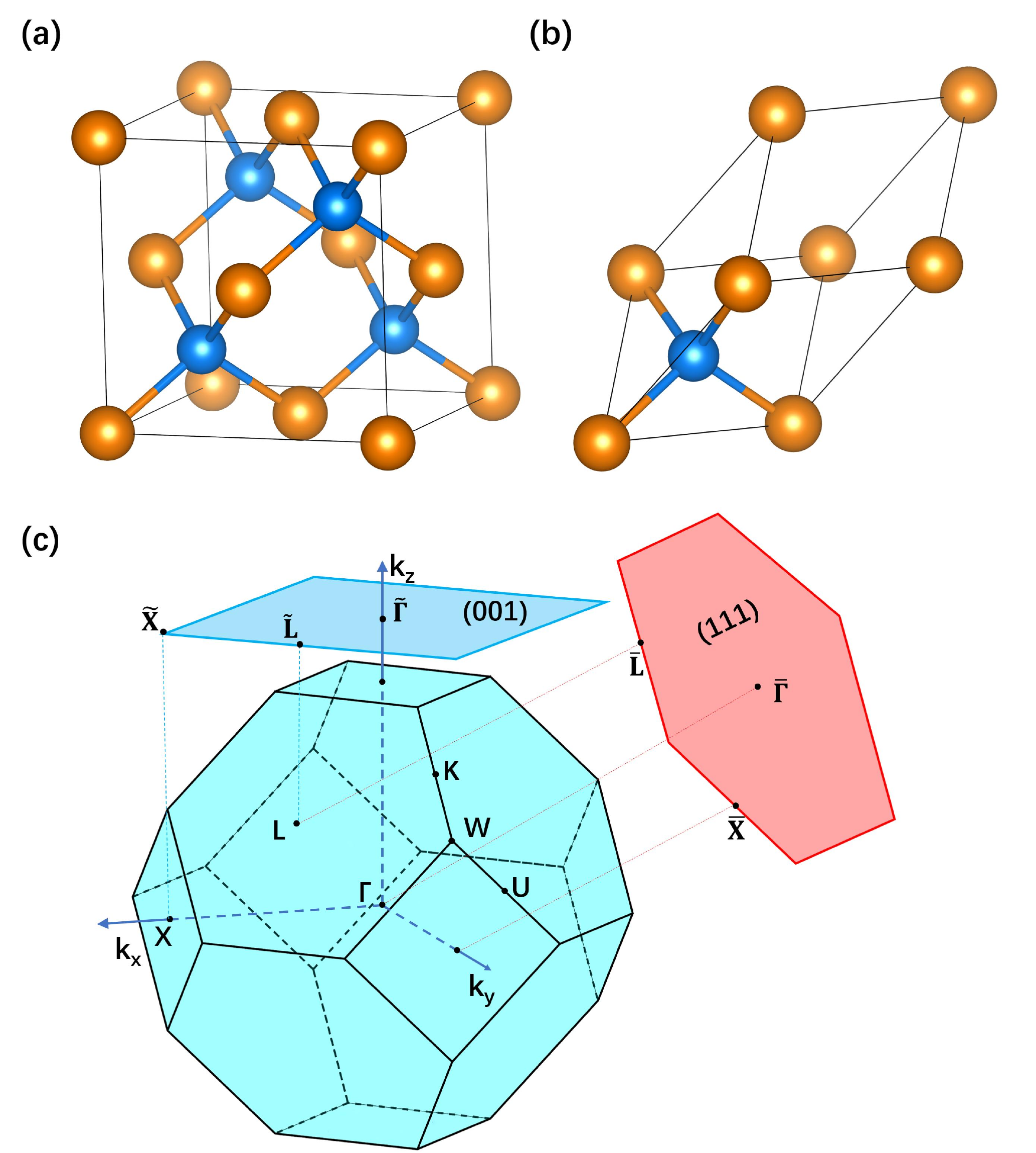}
	\caption{The crystalline structure of CdTe and its BZ. (a) The conventional and (b) primitive unit cells of CdTe with the space group $F$-$43m$ (No. 216). The Cd and Te atoms are represented by blue and orange spheres, respectively. (c) The fcc BZ and the corresponding (001) and (111) surface BZs.
\label{figure1}}
\end{figure}

We performed first-principles calculations using Vienna $ab$ $initio$ simulation package \cite{Kresse2} within the framework of density-functional theory \cite{Kohn}. The generalized gradient approximation with the Perdew-Burke-Ernzerhof (PBE) formalism  was selected to describe the exchange-correlation functional \cite{Perdew1,Perdew2}. The interactions between ions and valence electrons were treated by projector augmented wave method \cite{Kresse4,Ceperley1980} with the plane-wave cutoff energy of 400 eV. The full Brillouin-zone (BZ) was sampled by 21$\times$21$\times$21 Monkhorst-Pack grid \cite{Monkhorst} for the primitive unit cell. The crystal structure was fully relaxed until the forces on each atom were smaller than 0.001 eV/{\AA}. We carried out lattice dynamics calculations using the finite displacement method to generate real-space interatomic force constants within a 4$\times$4$\times$4 supercell. As CdTe is a typical polar material, we further considered the non-analytical term correction added in the dynamical matrix to avoid the degeneracy of transverse optical and longitudinal optical branches as $\mathbf{q}\rightarrow 0$ \cite{PhysRevB.1.910}. The Born effective charge tensors and static dielectric constant tensors were obtained from density functional perturbation theory. The phonon dispersions were further derived by diagonalizing the lattice dynamical matrix within the PHONOPY package \cite{Togo2015}.  We also calculated phonon spectra of CdTe using different functionals [detailed results can be found in Supplemental Material (SM)  \cite{SM}], and we confirmed the topological phonon features of CdTe. To reveal topological features of type-II Weyl phonons in CdTe, we constructed a Wannier tight-binding (TB) Hamiltonian of phonons from real-space interatomic force constants \cite{WU2017}.

\begin{figure}
	\centering
	\includegraphics[scale=0.29]{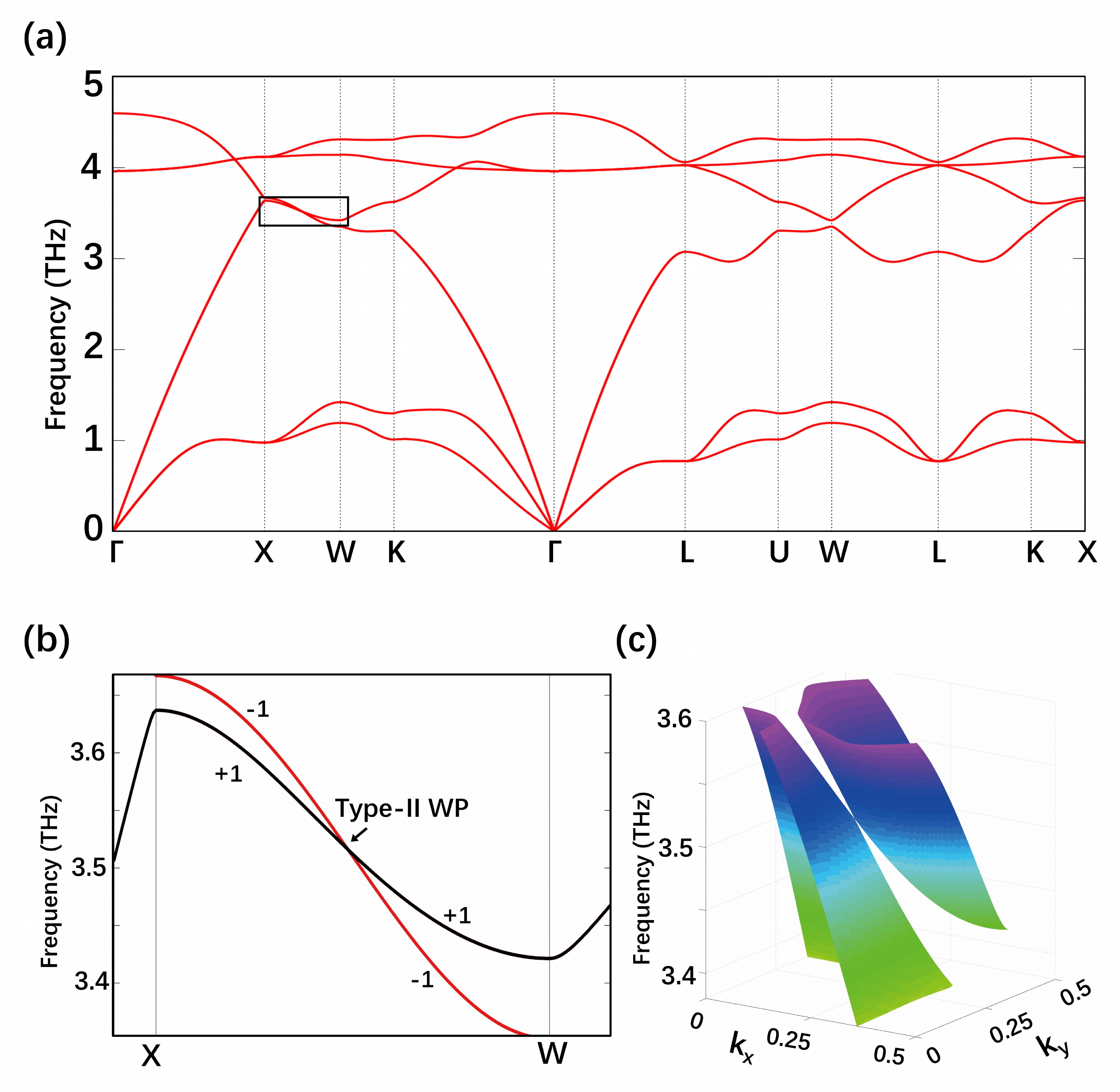}
	\caption{(a) The phonon spectra of CdTe. (b) The zoom-in phonon spectra marked by a black box in (a). The two inverted crossing branches belong to opposite eigenvalues $\pm 1$ of $C_2$. (c) The inverted phonon branches form a titled WP in the $q_x$-$q_y$ plane with $q_z = 2\pi/a$ of the BZ.
\label{figure2}}
\end{figure}

As shown in Fig. \ref{figure1}(a), zinc-blende CdTe crystallizes in a face-centered cubic (fcc) lattice with a lattice constant $a=6.60$ {\AA}, which agrees well with the experimental value $a=6.48$ {\AA} \cite{ref1}. There are one Cd atom and one Te atom in the fcc primitive unit cell, which occupy the (0, 0, 0) site and (${1}/{4}$, ${1}/{4}$, ${1}/{4}$) site  [see Fig. \ref{figure1}(b)], respectively.  CdTe is non-centrosymmetric with breaking the inversion symmetry. The fcc BZ and the corresponding (100) and (111) surface BZs are given in Fig. \ref{figure1}(c).

The calculated phonon spectra of CdTe with non-analytical term correction along the high-symmetry lines are shown in Fig. \ref{figure2}(a),
which match well with previous theoretical and experimental results \cite{Rowe1974, PhysRevB.56.8691, PhysRevB.53.9052, PhysRevB.47.3588}. It is clear to see that there is no acoustic-optical branch gap [see Fig. \ref{figure2}(a)]. From the enlarged view of phonon dispersion in Fig. \ref{figure2}(b), we find that a double-degenerate point is present in the $X$-$W$ direction. The degenerate point originates from the crossing of the longitudinal acoustic and the transverse optical branches at the frequency of $\omega_{\mathrm{wp}}= 3.5$ THz. Accordingly, this fact corresponds to the appearance of two phonon branches inverted at the $X$ point.  Beyond the PBE functional, we also calculate the phonon spectra of CdTe using different functionals, and the results show that the phonon branch inversion is robust \cite{SM}. Since the frequency of the acoustic branch is often lower than that of the optical branch, the presence of phonon branch inversion indicates that unusual features may occur in CdTe. We further check the little group along the $X$-$W$ line, and find that the crossing with linear dispersion along this direction is with respect to the 2-fold rotational symmetry $C_2$ at the boundary of the fcc BZ. The two crossing branches belong to opposite eigenvalues $\pm 1$ of $C_2$. As shown in Fig. \ref{figure2}(c), we illustrate the phonon dispersion around the crossing point in the $q_x$-$q_y$ plane with $q_z = 2\pi/a$. The figure shows that the crossing of two inverted branches forms a titled Dirac point, indicating that elementary excitations near the crossing point may belong to type-II Weyl phonons.

\begin{figure}
	\centering
	\includegraphics[scale=0.295]{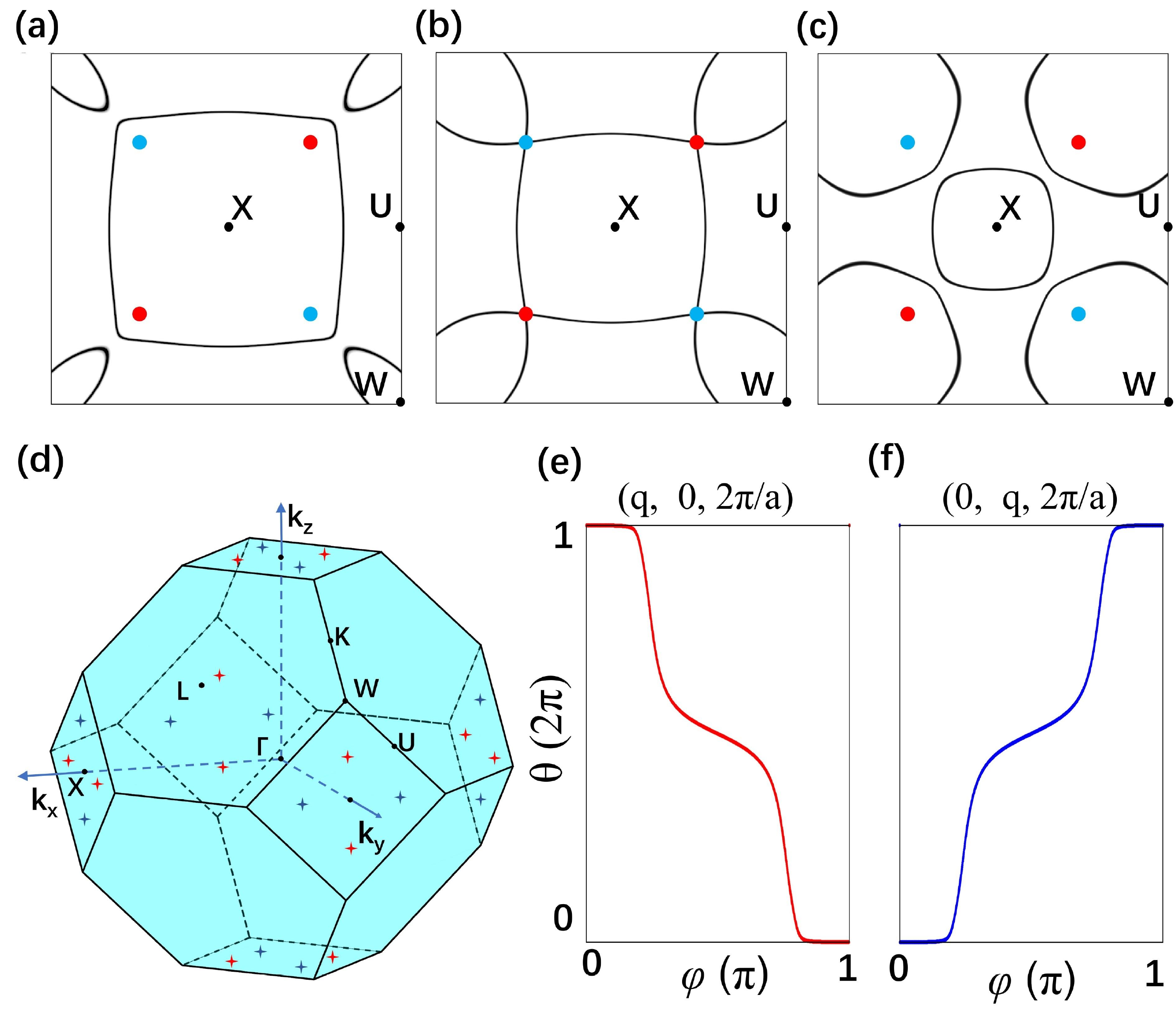}
	\caption{The cross-section of the bulk iso-frequency surface in the $q_x$-$q_y$ plane with $q_z = 2\pi/a$ with the frequency (a) below, (b) at, and (c) above  $\omega_{\mathrm{wp}}$. (d) The distribution of WPs at the boundaries of BZ, i.e., the $q_x= 2\pi/a$, $q_y= 2\pi/a$, and $q_z= 2\pi/a$ planes.  The WPs with opposite chirality are marked as red ($\mathcal{C}=-1$) and blue ($\mathcal{C}=+1$) crosses, respectively.  The evolution of the average position of Wannier centers for a WP with (e) negative or (f)  positive chirality.
\label{figure3}}
\end{figure}

We further examine the iso-frequency surface to confirm that Weyl phonons in CdTe are type -II. The cross-section of the bulk iso-frequency surface in the $q_x$-$q_y$ plane with $q_z = 2\pi/a$ is illustrated in Figs. \ref{figure3}(a), \ref{figure3}(b), and \ref{figure3}(c). The iso-frequency surface consists of one acoustic and four optical pockets. We can see that the acoustic and optical pockets touch at the WPs when $\omega = \omega_{\mathrm{wp}}$ [see Fig. \ref{figure3}(b)], exhibiting open iso-frequency contours around each WP. Therefore, CdTe hosts type-II Weyl phonons as its inversion symmetry is broken.  Considering the crystal symmetry, there are a total of 12 such WPs that lie along the high-symmetry lines $X$-$W$ at the boundaries of the fcc BZ (i.e., $q_x= 2\pi/a$, $q_y= 2\pi/a$, and $q_z= 2\pi/a$ planes) [see Fig. \ref{figure3}(d)].

In order to determine the chirality of type-II Weyl phonons, we employ the Wilson-loop method within the evolution of the average position of Wannier centers \cite{WU2017, Yu2011} [see Figs. \ref{figure3}(e) and \ref{figure3}(f)]. Our results show that the adjacent WPs along the orthogonal high-symmetry line $X$-$W$ in the $q_x$-$q_y$ plane with $q_z = 2\pi/a$ have opposite Chern numbers, which are well separated in momentum space [see Fig. \ref{figure3}(d)]. The positions of these WPs can be expressed by a single parameter $\mathbf{q}_{\mathrm{wp}}$ with the chirality $\mathcal{C}=+1$ at $(2\pi/a, 0, \pm q)$, $(\pm q, 2\pi/a, 0)$, and $(0, \pm q,  2\pi/a)$, and with chirality $\mathcal{C}=-1$ at $(2\pi/a, \pm q, 0)$, $(0, 2\pi/a, \pm q)$, and $(\pm q, 0,  2\pi/a)$, where $q=0.054$ {\AA}$^{-1}$. When the phonon frequency is tuned from below [see Fig. \ref{figure3}(a)] to above [Fig. \ref{figure3}(c)]  $\omega_{\mathrm{wp}}$, all the WPs are inside from acoustic to optical pockets.  The total Chern number in either acoustic or optical pockets is zero since each type of pockets encloses an equal number of WPs with opposite chirality.

Next, we carry out an extensive symmetry analysis to understand the nature of the symmetry-protected type-II Weyl phonons in CdTe. The nonmagnetic space group $F$-$43m$ contains three 2-fold rotational symmetries $C_2 ^{i}$ ($i=x$, $y$, or $z$). In addition, the time-reversal ($\mathcal{T}$) symmetry of a phonon system is always conserved.  In the following, we show that the type-II WPs appear at the boundaries of the fcc BZ is due to the coexistence of $C_2$ and $\mathcal{T}$. Generally, we use a $2\times 2$ $\mathbf{k}\cdot \mathbf{p}$ Hamiltonian to describe the two crossing branches of phonons,
\begin{equation}\label{Hamiltonian}
H(\mathbf{q})=d_{x}(\mathbf{q})\sigma_{x}+d_{y}(\mathbf{q})\sigma_{y}+d_{z}(\mathbf{q})\sigma_{z},
\end{equation}
where $H$ is referenced to $\hbar \omega_{\mathrm{wp}}$ with the Planck constant $\hbar$, $d_{x,y,z} (\mathbf{q})$ are real functions with wavevector $\mathbf{q}=$($q_x$, $q_y$, $q_z$), and $\sigma_{x,y,z}$ are the corresponding Pauli matrices.

We first consider $C_2^{z}$ with the rotational axis along $z$. As the two crossing branches belong to opposite eigenvalues of $C_2$, the product operator $C_2^{z}\mathcal{T}$ can be represented by $C_2^{z}\mathcal{T} =\sigma_z K$, where $K$ is the complex conjugate operator. The symmetry $C_2^{z}\mathcal{T}$ requires\begin{equation}
H(-C_2^{z}\mathbf{q})=C_2^{z}\mathcal{T}H(\mathbf{q})\mathcal{T}^{-1}{C_2^{z}}^{-1},
\end{equation}
which gives
\begin{equation}\label{C2z}
\begin{split}
&d_{x}(q_{x},q_{y},q_{z})=-d_{x}(q_{x},q_{y},-q_{z}), \\
&d_{y,z}(q_{x},q_{y},q_{z})= d_{y,z}(q_{x},q_{y},-q_{z}).
\end{split}
\end{equation}
Considering the periodic condition of $H(\mathbf{q})$, Eq. (\ref{C2z}) requires $q_z = {2\pi}n/{a}$ ($n\in \mathbb{Z}$). This implies that the crossing points may be present at the boundary of the fcc BZ with $q_z = {2\pi}/{a}$. In this case, $d_{x}(q_x, q_y, 2\pi/a)\equiv 0$. Besides, the extra  $C_2^{x}$ with the rotational axis along $x$ or $C_2^{y}$ with the rotational axis along $y$ further constrains $H(q_x, q_y, {2\pi}/{a})$, which leads to
\begin{equation}\label{C2x}
d_{y,z}(q_{x},q_{y},\frac{2\pi}{a})= -d_{y,z}(q_{x},-q_{y},\frac{2\pi}{a}),
\end{equation}
or
\begin{equation}\label{C2y}
d_{y,z}(q_{x},q_{y},\frac{2\pi}{a})= -d_{y,z}(-q_{x},q_{y},\frac{2\pi}{a}).
\end{equation}

As a result, Eqs. (\ref{C2x}) and (\ref{C2y}) force the crossing points  to be along the $q_x$ or $q_y$ axis, which are protected by the $C_2^{x}$ or $C_2^{y}$ symmetry. That is to say, the coexistence of crystal and $\mathcal{T}$ symmetries guarantees that the type-II WPs are located along the high-symmetry lines $X$-$W$ in the plane with $q_z = {2\pi}/{a}$. Through the similar analysis, the combination of $C_2$ and $\mathcal{T}$ can also derive that the WPs can be located along the high-symmetry lines $X$-$W$  in the $q_x = {2\pi}/{a}$ or $q_y = {2\pi}/{a}$ plane. Because the high-symmetry lines $X$-$W_{x,y,z}$ are perpendicular to each other, the WPs with opposite chirality in CdTe are well separated. Unlike the Weyl fermions, these phonon WPs at the boundaries of the fcc BZ are always symmetry-protected due to the absence of SOC in phonon systems, resulting in the robust protected feature of type-II Weyl phonons.

\begin{figure}
	\centering
	\includegraphics[scale=0.19]{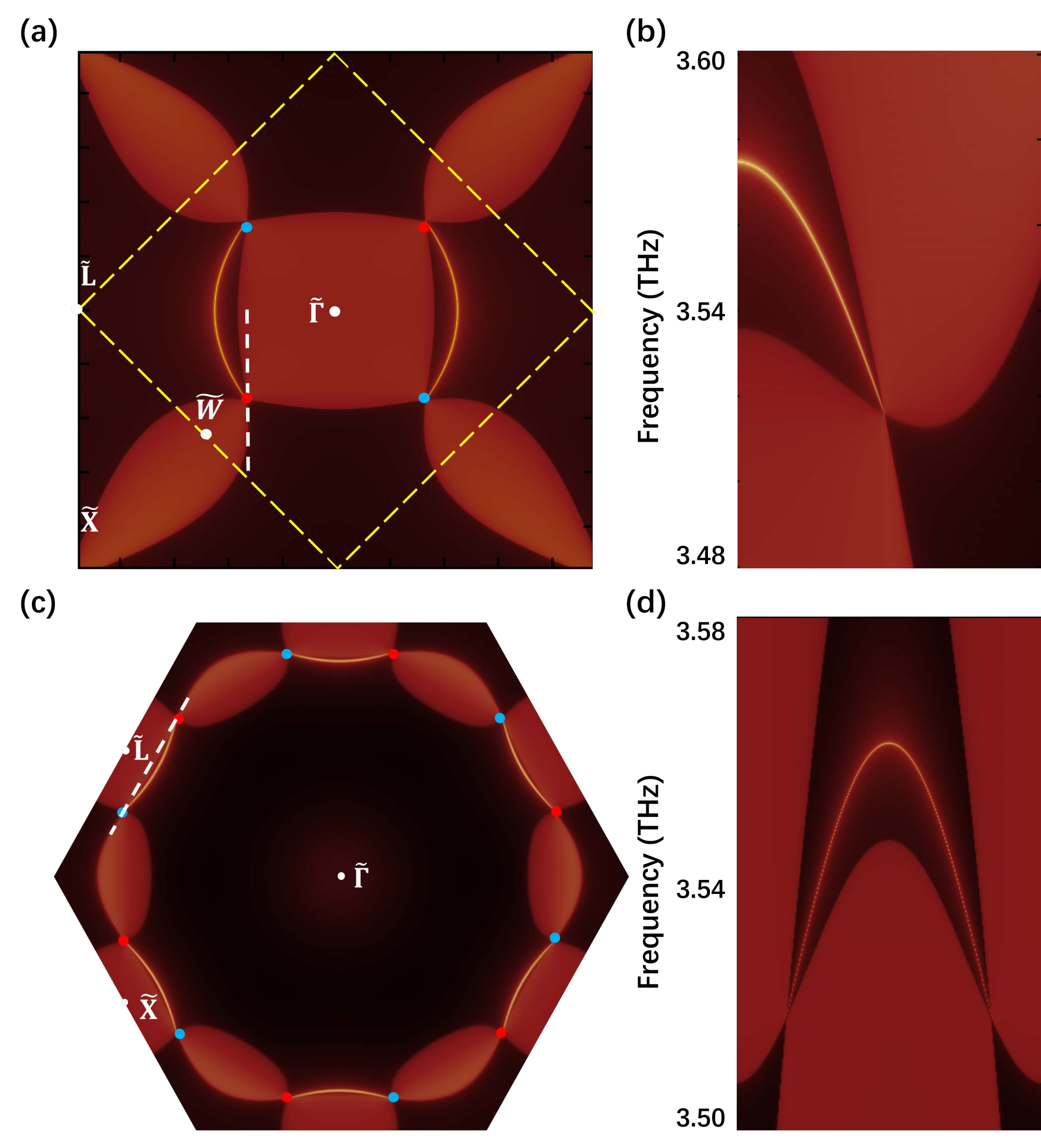}
	\caption{The phonon surface states of CdTe.  (a) The phonon surface arcs projected on the semi-infinite (001) surface. Region inside the yellow-dashed lines denotes the part of (001) surface BZ projected from the $q_z = 2\pi/a$ plane. (b) The LDOS for the (001) surface along the white line marked in (a). (c) The phonon surface arcs projected on the semi-infinite (111) surface. (d) The LDOS for the (111) surface along the white line marked in (c). The blue and red dots indicate the projections of WPs with $\mathcal{C}= +1$ and $\mathcal{C}=-1$, respectively.
\label{figure4}}
\end{figure}

To proceed our illustration on topologically protected features of type-II Weyl phonons in CdTe, we calculate the phonon surface states using the iterative Green's function method  \cite{Sancho1984}  with a phonon TB Hamiltonian as implemented in the WANNIERTOOLS package \cite{WU2017}. The calculated phonon local density of states (LDOS) and surface arcs for the semi-infinite (001) and (111) surfaces are shown in Fig. \ref{figure4}. For the semi-infinite (001) surface, the four WPs in the $q_x$-$q_y$ plane with $q_z = 2\pi/a$ are projected to the diagonal lines (i.e., $\widetilde{\Gamma}$-$\widetilde{W}$-$\widetilde{X}$) of the (001) surface BZ [see Fig. \ref{figure4}(a)]. As expected, this iso-frequency surface shows that there are  two long surface arcs connecting projections of two WPs with opposite chirality. These phonon surface arcs with respect to the 2-fold rotational symmetry $C_2^z$ along the $z$ axis are  clearly visible. We also plot the corresponding LDOS in Fig. \ref{figure4}(b). It is found that the phonon surface states are terminated at the projection of titled Dirac cone, further confirming the nontrivial features of type-II Weyl phonons. The iso-frequency surface for the (111) surface is shown in Fig. \ref{figure4}(c). Each surface arc starts and ends at projections of a pair of WPs with opposite Chern numbers. We can see that there are six symmetry-distributed surface phonon arcs due to the hexagonal symmetry of the (111) surface BZ [see Fig. \ref{figure4}(c)]. Similarly, the corresponding LDOS for the CdTe (111) surface are shown in Fig. \ref{figure4}(d).

In conclusion,  by performing first-principles calculations and symmetry analysis, we identify that ideal type-II Weyl phonons exist in a well-known semiconducting material CdTe, which originate from the phonon branch inversion between the longitudinal acoustic and transverse optical branches. The symmetry arguments indicate that the type-II WPs in CdTe are protected by time-reversal $\mathcal{T}$ and 2-fold rotational symmetries $C_2$. Due to the lack of SOC, the WPs exactly are located along the high-symmetry lines at the boundaries of the fcc BZ, leading to robust topological phonon features. The phonon surface arcs connecting the WPs with opposite chirality are guaranteed to be very long and are easily detectable in experiments. These extremely long phonon arcs can provide a robust one-way phonon propagation channel, and the elastic waves of topological surface phonon states can only propagate on the surface of a material in a certain direction without backscattering from imperfections. The ideal topological features of symmetry-protected type-II Weyl phonons in CdTe is useful in future applications of phonon transport, including the fabrication of topological thermal devices.


~~~\\

This work was supported by the National Natural Science Foundation of China (NSFC, Grants No. 11674148, No. 11334003, No. 91634106, No. 11404159, and No. 11847301), the Guangdong Natural Science Funds for Distinguished Young Scholars (No. 2017B030306008),the Fundamental Research Funds for the Central Universities of China (Grants No. cqu2018CDHB1B01,No. 2019CDXYWL0029, and No. 2019CDJDWL0005), and the Science, Technology and Innovation Commission of Shenzhen Municipality (No. ZDSYS20170303165926217).\\

B.W.X. and R.W. equally contributed to this work.


\end{document}